\documentclass[letterpaper, 10 pt, conference]{ieeeconf}  

\IEEEoverridecommandlockouts                              

\usepackage{balance}
\let\labelindent\relax
\usepackage{enumitem}
\usepackage{cite}
\usepackage{amsmath,amssymb,amsfonts}
\usepackage{algorithmic}
\usepackage{graphicx}
\usepackage{textcomp}
\usepackage{xcolor}
\newtheorem{theorem}{\textbf{Theorem}}[section]

\newtheorem{definition}[theorem]{\textbf{Definition}}
\newtheorem{assumption}[theorem]{\textbf{Assumption}}
\newtheorem{proposition}[theorem]{\textbf{Proposition}}
\newtheorem{remark}[theorem]{\textbf{Remark}}
\usepackage[colorinlistoftodos]{todonotes}
\usepackage{color}
\usepackage{tcolorbox}

\def\BibTeX{{\rm B\kern-.05em{\sc i\kern-.025em b}\kern-.08em
    T\kern-.1667em\lower.7ex\hbox{E}\kern-.125emX}}

\title{\LARGE \bf
Safe Control Design through Risk-Tunable Control Barrier Functions
}

\author{Vipul K. Sharma and S. Sivaranjani
\thanks{The authors are with the School of Industrial Engineering at Purdue University, West Lafayette, IN 47907, USA.
        {\tt\small \{sharm697,sseetha\}@purdue.edu}}
}

\begin{document}

\maketitle
\thispagestyle{empty}
\pagestyle{empty}


\begin{abstract}
We consider the problem of designing controllers to guarantee safety for a class of nonlinear systems under uncertainties in the system dynamics and/or the environment. We define a class of uncertain control barrier functions (CBFs), and formulate the safe control design problem as a  chance-constrained optimization problem with uncertain CBF constraints. We leverage the scenario approach for chance-constrained optimization to develop a risk-tunable control design that provably guarantees the satisfaction of uncertain CBF safety constraints up to a user-defined probabilistic risk bound, and provides a trade-off between the sample complexity and risk tolerance. We demonstrate  the performance of this approach through simulations on a quadcopter navigation problem with obstacle avoidance constraints. 
\end{abstract}

\section{INTRODUCTION}
Safety is a central consideration in the design of  autonomous systems  that operate in uncertain and unknown environments, spanning numerous applications such as automated driving, robotics, and unmanned aerial vehicles.  The problem of designing controllers that provably guarantee hard constraints on the safety of autonomous systems is a long-studied topic, and has seen a recent resurgence in interest due to advances in learning-based approaches as well as emerging applications such as self-driving cars, where autonomous systems are expected to closely interact with humans in safety-critical settings.

Model-based design approaches to guarantee safety often involve imposing Control Barrier Functions (CBF) constraints \cite{ames2016control} on the control design problem. CBF constraints employ a Lyapunov-like argument to guarantee the invariance of a desired `safe set' under the designed control law, essentially guaranteeing that a system that starts in a safe set  always remains in the safe set. We refer the reader to \cite{ames2019control} for a comprehensive survey on the various classes of CBF conditions commonly employed in control design. In situations where the control design problem involves uncertainty in the system dynamics or environment, robust control invariance conditions of a similar nature may be enforced to guarantee safety of the control design \cite{choi2021robust,buch2021robust,sadraddini2017provably,gurriet2018towards,cheng2020safe,cosner2021measurement,santillo2021collision}. 

While CBF-based designs, including their robust versions, are effective in guaranteeing safety, a key challenge is that they are often conservative, imposing robustness to worst-case uncertainties, which may themselves be difficult to characterize in dynamic environments. Further, they may come at a great efficiency cost (both in terms of time-efficiency and control cost) that can make autonomous task execution  practically unviable in several applications. Finally, notions of safety and risk can vary widely by domain. For
example, in some applications, small safety constraint violations may be tolerable in order to increase task efficiency. Therefore, it is desirable to introduce probabilistic notions of safety with risk-efficiency tradeoffs that can be selected by designers based on application-specific considerations. In this context, this paper addresses the problem of introducing the notion of tunable risk into the safe control design problem. 

Specifically, we consider a class of nonlinear control-affine systems with an additive uncertainty that may arise due to unknown dynamics and/or environmental variables (including obstacles). For this class of systems, we formulate a safe control design problem with  uncertain CBF constraints that must be satisfied with a user-defined probabilistic risk bound. We pose this problem in a chance-constrained optimization setting, and propose a sampling-based control design based on the scenario approach \cite{campi2009scenario,calafiore2006scenario, campi2018introduction} that allows the designer to tune the risk bound for safety constraint satisfaction, and provides a trade-off between the risk bound and the sample complexity of the problem. We demonstrate the performance of this design by simulation on a quadcopter navigation problem with obstacle avoidance constraints. 

\subsection{Related Work and Contributions}
A common approach to safe control design for systems with uncertainties involves modeling the uncertainty by a known process, with Gaussian Process (GP) models receiving significant attention in typical MPC based control designs \cite{hewing2019cautious,bradford2018stochastic}, as well as CBF-based designs for safety-critical systems \cite{cheng2019end,cheng2020safe,hu2023safe,luo2022sample}. However, these works  do not typically consider the probabilistic notions of safety required to incorporate risk tunability that is the subject of this work. 
Control designs incorporating probabilistic notions of safety through chance-constrained CBFs have recently been proposed \cite{wang2021chance,blackmore2006probabilistic,nakka2020chance,salehi2022learning,khojasteh2020probabilistic,luo2022sample}. In general, such chance-constrained optimization problems are non-convex, even when the original CBF constraints are convex, and are often NP-hard \cite{ben1998robust,ben2002tractable}.  Solutions to such probabalistic safe control design problems typically involve either approximating the uncertainty by GP (or similar) models \cite{salehi2022learning,nakka2020chance,khojasteh2020probabilistic}, or deriving convex relaxations or over-approximations that make the problem tractable \cite{wang2021chance,blackmore2006probabilistic}.  However, GP models may not hold in several safety-critical applications, and convex over-approximations may lead to conservative designs. In this paper, we introduce a sampling-based safe control design framework based on the scenario-approach for chance-constrained optimization \cite{calafiore2006scenario} that does not make any assumptions regarding the underlying distribution of the uncertainty. The  scenario approach has recently been utilized for safety verification with CBFs \cite{akella2022barrier}; however, this work does not address control design, which is the central problem considered in this paper. 

In this landscape, the key contribution of this paper is to introduce a framework to design controllers that can guarantee probabilistic notions of safety with user-defined risk bounds, with the advantage that the risk-efficiency and sample complexity trade-offs can be tuned by designers based on domain-specific requirements. We also note that most of the above safe control designs employ projection-based approaches, where a baseline controller (that is designed based on a separate optimization problem or is assumed to be given) is  minimally modified through CBF constraints to enforce safety. Such projection-based approaches may result in sub-optimal controllers. In contrast, our approach directly solves a chance-constrained problem to optimize performance metrics while simultaneously guaranteeing safety, without the need for a two-stage solution.

\subsection{Organization}
This paper is organized as follows. Section \ref{sec:problem} formulates the safe control design problem under uncertainties as a chance-constrained optimization problem with control barrier function constraints. Section \ref{sec:control} provides a risk-tunable design based on the scenario approach to solve the safe control design problem. Section \ref{sec:case} demonstrates the design through simulation on a quadcopter navigation problem with uncertain obstacles. The proofs of all the results in this paper are presented in the Appendix. 

\subsection{Notation}
We denote the sets of real numbers, positive real numbers including zero, and $n$-dimensional real vectors by $\mathbb{R}$, $\mathbb{R}_{+}$ and $\mathbb{R}^{n}$ respectively. 
	For a matrix $A \in \mathbb{R}^{m\times n}$, $A^T\in \mathbb{R}^{n \times m}$ represents its transpose. 	A symmetric positive definite matrix $P \in \mathbb{R}^{n \times n}$ is represented as $P>0$ (and as $P\geq 0$, if it is positive semi-definite). The standard identity matrix is denoted by $I$, with dimensions clear from the context. An $(n\times m)$ matrix  with all elements equal to 1 is denoted by $\mathbf{1}_{n\times m}$. Similarly, an $(n\times m)$ matrix  with all elements equal to  zero is denoted by $\mathbf{0}_{n\times m}.$ 
For any $N_1, N_2 \in \mathbb{R}_{+}, \begin{pmatrix}
        N_1 \\ N_2
    \end{pmatrix}$  represents the number of ways to choose $N_2$ items from a set of $N_1$ items. 
    
\section{PROBLEM FORMULATION}\label{sec:problem}
We begin by formulating the safe control design problem addressed in this paper.
We consider a nonlinear dynamical system with control-affine dynamics given by,
\begin{equation}\label{dynamics}
x_{t+1} = f(x_t) +g(x_t)u_t+d_t,
\end{equation}
where $x_t \in X \subset \mathbb{R}^n$ is the state,  $u_t \in U \subset \mathbb{R}^m$ is the control input, and $d_t \in U \subset \mathbb{R}^n$ is an additive disturbance at time $t\in \mathbb{R}_{+}$, and  $f:\mathbb{R}^n\to \mathbb{R}^n$ and $g:\mathbb{R}^n\to \mathbb{R}^n\times \mathbb{R}^m$ are locally Lipschitz continuous. Moreover, we suppose that $U=[u^{l},u^{h}]$, where $u^{l}$ and $u^{h}$ are actuator constraint lower and upper bounds respectively. 

\begin{assumption}
    The dynamical system is assumed to be forward complete, that is, the solution to \eqref{dynamics} is defined for all initial conditions $x_0 \in X$ and all admissible control inputs $u_t \in U$ for all time $t \in \mathbb{R}_{+}$.
\end{assumption}

We begin by defining control barrier functions (CBFs) and associated conditions that we will utilize in formulating the safe control design problem that is the subject of this work.

%
%

\subsection{Control Barrier Functions (CBF)}
Let $\mathcal{S}$ be a safe set, defined by the super-level set of a continuously differentiable function $\bar h \colon X \rightarrow \mathbb{R}$ as,
\begin{equation}\label{safe_set}
    \mathcal{S} = \{x \in X \colon \bar h(x) \geq 0\}.
\end{equation}


\noindent If the states always remain within this set, 
then we can guarantee the safety of the system as follows. 


\begin{definition}
   The dynamical system \eqref{dynamics} is said to be \underline{safe} with respect to set $\mathcal{S}$ if $\mathcal{S}$ is \textit{forward-invariant}, that is, $\forall x_0 \in \mathcal{S}$, $x_t \in \mathcal{S}, \forall t \in \mathbb{R}_{+}$. 
\end{definition}

A standard approach to establish forward invariance of the safe set $\mathcal{S}$ is to derive sufficient conditions using a  Lyapunov-like argument  as follows. 
\begin{theorem}[Adapted from \cite{agrawal2017discrete}]\label{thm:cbc}
  A continuously differentiable function 
$\bar h \colon X \rightarrow \mathbb{R}$ is a control barrier function for dynamical system \eqref{dynamics} and renders the set $\mathcal{S}$ safe if
there exists a control input $u_t$ and a constant $\eta \in [0, 1]$ such that for all 
$x_t \in \mathcal{S}$, we have
\vspace{-1mm}
\begin{equation} \label{eq:cbf_ineq1}
    \bar h(x_{t+1}) - (1-\eta)\bar h(x_t)
    \geq 0.
\end{equation}
\end{theorem}

In Theorem \ref{thm:cbc}, the constant $\eta$ determines how strongly the CBF pushes the states into the safe set \cite{cheng2019end}. 

\subsection{Uncertain Control Barrier Functions}
With the safe set defined in \eqref{safe_set} and the CBF condition defined in Theorem \ref{thm:cbc}, we now focus on the scenario where there is uncertainty in the CBF condition \eqref{eq:cbf_ineq1}, either due to partially known/uncertain dynamics or due to the operating environment. We define a robust CBF condition as follows.

\begin{theorem}\label{thm:cbc_uncertain}
 The dynamical system \eqref{dynamics} can be rendered safe with safe set $\mathcal{S}$ if, for all  $d_t\in D$, there exists control input $u_t \in U$ and a constant $\eta \in (0, 1]$ such that
 \vspace{-1mm}
 {\begin{equation} \label{eq:cbf_ineq}
 \vspace{-0.5mm}
    L(x_t,u_t, d_t) :=  -h(x_t,u_t,d_t)-\eta\bar h(x_t)\leq 0, \vspace{-0.3mm}
\end{equation}}
where $h(x_t,u_t,d_t):=\bar h(x_{t+1}))-\bar h(x_t)$.
\end{theorem}




\begin{remark}\label{rem:cbf_uncertainty} The CBF condition in Theorem \ref{thm:cbc} can be appropriately defined to capture commonly encountered uncertainties in system dynamics and operation as follows:
\begin{itemize}[leftmargin=*]
    \item \textit{{Example E1 - Uncertainty in System Dynamics or Environment:}} Consider the case where part of the system dynamics is unknown, that is,  $d_t \in D$ is the unknown part of the dynamics in \eqref{dynamics}.  A candidate CBF for such a case is an affine CBF  $\bar h(x):=p^Tx+q,$
    where $p \in \mathbb{R}^n$ and $q \in \mathbb{R}$. Then, the CBF condition in \eqref{eq:cbf_ineq} for this case can be written as $L(x_t,u_t, d_t) =-[p^T(f(x_t)+g(x_t)u_t+d_t) + q]  \nonumber+ (1-\eta)(p^T x_t +q) \leq 0.$ 
    An identical CBF condition holds when $d_t$ represents an additive exogenous disturbance arising from the interaction of the system with the environment.
    \item \textit{Example E2 - Obstacle Avoidance:}  Consider a navigation problem with obstacle avoidance constraints, where the objective is to maintain a safe distance between an agent (such as a mobile robot or aerial vehicle) and an obstacle with position $x^{obs}_t \in \mathbb{R}^n$ at time $t$. Let the obstacle position be stationary and uncertain, i.e. $x^{obs}_{t}=x^{o}_{t}+d_t$ and $x^{obs}_{t+1}=x^{obs}_{t}$, where $x^{o}_t$ is  known  and $d_t$ is  unknown.
    For this case, we may define $\bar h(x_t) := \|x_t-x^{obs}_t\|^2_{2}-r_s$, where $r_s$ is a user-defined safety margin. Then, the CBF condition in \eqref{eq:cbf_ineq} can be written as 
                $L( x_t, u_t, d_t) =-[\| f(x_t)+ g(x_t) u_t- x^{o}_t-d_t\|^2_2  -r_s ]  
                    +(1-\eta)\|x_t-x^{o}_t-d_t\|^2_{2}-r_s
                    \leq 0. $
\end{itemize}
\end{remark}       
\vspace{-0mm}
\subsection{Control Design Problem}
With this uncertain CBF formulation, the goal is to design control inputs $u_t$ that render the system \eqref{dynamics} safe $\forall d\in D$. To obtain such control inputs at each time step $t$, we formulate a robust control design problem with constraints given by condition  \eqref{eq:cbf_ineq} in Theorem \ref{thm:cbc_uncertain}. Consider the robust control design problem (${RCP}^t$) at time step $t$ expressed as

\vspace{-3mm}
{\small
\begin{equation}\label{eq:rcp}
RCP^t:
    \begin{aligned}
        \underset{u_t}{\mathrm{min}}\ &C(u_t) \\
        \text{s.t.} \ \  &L(x_t,u_t,d_t) \leq 0, \ \forall d_t \in D, 
        u^{l} \leq u_t \leq u^{h},
    \end{aligned} 
\end{equation}}
where $C(u_t)$ is the cost function. 


There are several difficulties involved in solving the problem $RCP^t$. First, $RCP^t$ involves a possibly infinite number of constraints. Even if the problem is assumed to be convex, this class of problems is, in general, NP-hard \cite{ben1998robust,ben2002tractable,el1998robust}. One common approach is to consider the a `worst-case' solution to the {RCP}, where the CBF constraint in \eqref{eq:rcp} is replaced by $L(x_t,u_t,d_t^{*})\leq 0$, where $d_t^{*}=\max \{  d : d \in D\}$. However, such a design would be overly conservative in many applications, and result in decreased performance metrics such as time-efficiency or control effort. Further, in many applications, a small tolerance towards risk is generally acceptable during operation under uncertainty. 





We quantify the  risk-tolerance  in the safe control design problem in terms of violation probability of the CBF condition as follows.

\begin{definition}
    (Violation Probability)
The probability of violation under control input $u_t$ is defined as
\begin{equation}
    V(u_t) := Prob\{d_t \in D : L(x_t,u_t,d_t) > 0\}.
\end{equation}
\end{definition}
For a given control input $u_t$, the probability that this input violates the CBF constraint $L(x_t,u_t,d_t) \leq 0$ is given by $V(u_t)$. Assuming a uniform probability density, the violation probability can be interpreted as a measure of the volume of `unsafe' uncertainty parameters $d_t$ such that the CBF constraint is violated. 

Now, select a \underline{tunable user-defined risk bound} $\epsilon \in (0,1)$ that quantifies the acceptable violation probability. Note that $\epsilon$ can be selected by the designer based on the application. Then, we define an $\epsilon$-level solution as follows:
\begin{definition}
     ($\epsilon$-level solution)
We say that $u_t \in U$ is an $\epsilon$-level solution, if $V(u_t) \leq \epsilon$, $\epsilon \in (0,1)$. 
\end{definition}

With these definitions, we now reformulate the RCP into a chance constrained problem (\textbf{$CCP^t(\epsilon)$}) as follows:

{\vspace{-5mm}
\begin{equation*}
CCP^t(\epsilon):
    \begin{aligned}
        \underset{u_t}{\mathrm{min}}\ &C(u_t) \\
        \text{s.t.} \ \  &Prob\{d_t \in D : L(x_t,u_t,d_t) \leq 0\} > 1-\epsilon, \\
        &u^{l} \leq u_t \leq u^{h}
    \end{aligned} 
\end{equation*}}

\begin{definition}($\epsilon$-safety)\label{def:epsilon-safe}
   The dynamical system \eqref{dynamics} is said to be \underline{$\epsilon$-safe} if for all $d_t \in D$, there exists a control input $u_t \in U$ solving the problem $CCP^t_{\epsilon}$. 
\end{definition}

We call the problem of designing a  control input $u_t$ that solves  this chance-constrained problem 
as the ``risk-tunable ''control design problem, and formally state it as follows. 

\textbf{{Risk-Tunable Control Design Problem}} $\mathcal{P}_r:$ Given a user-defined risk tolerance bound $\epsilon$, find control input $u_t$ solving $CCP^t(\epsilon)$ such that the dynamical system \eqref{dynamics} is rendered $\epsilon$-safe at every time $t$.




\section{RISK-TUNABLE CONTROL DESIGN}\label{sec:control}
In this section, we develop an approach to  solve the risk-tunable control design problem $\mathcal{P}_r$. We begin by making the following assumptions regarding the convexity of the chance-constrained problem $CCP^t_\epsilon$.

\begin{assumption} \label{assum:convex}
   \begin{itemize}
       \item[(i)] We suppose that the objective function $C(u_t)$ is a convex function in the control input $u_t$. 
       \item[(ii)] Let $u_t \subset U$ be a convex and closed set, and let $D \subset \mathbb{R}^n$.
We assume that $L(x_t,u_t, d_t) :  U \times D \rightarrow (-\infty, \infty]$ is continuous and convex in $U$, for any $d_t \in D$.
   \end{itemize}
\end{assumption}

\begin{remark} Note that an exact numerical solution of $CCP^t(\epsilon)$ is intractable, see \cite{prekopa2013stochastic,vajda2014probabilistic}. Moreover, $CCP^t(\epsilon)$ is in general non-convex, even when Assumption \ref{assum:convex} holds.
\end{remark}

There are several ways to solve such chance-constrained problems, including approximating the uncertainty by a known process (e.g., Gaussian process) \cite{hewing2019cautious,bradford2018stochastic}, developing convex relaxations or approximations \cite{nemirovski2007convex,blackmore2006probabilistic}, and sampling-based approaches \cite{calafiore2006scenario,campi2009scenario}.  In this work, we develop a sampling-based design based on the scenario approach \cite{calafiore2006scenario}. The key idea is that if a sufficient number of samples of the uncertainty $d_t \in D$ can be extracted, then we can obtain a controller that renders the system safe for most  uncertainties up to a risk tolerance threshold.

\begin{definition}
    (\textbf{Scenario Design}) 
Assume that $N$ independent identically distributed samples $d^1_t, ..., d^N_t$ are drawn according to probability $Prob$. A scenario design problem is given by the convex optimization problem: 

\vspace{-3mm}
{\small
\begin{equation} \label{scn_des}
RCP^{t}_N:
    \begin{aligned}
        \underset{u_t}{\mathrm{min}}\, & C(u_t)
        \\
        \text{s.t.}   \ &L(x_t,u_t,d^i_t) \leq 0,  i \in     \{1,...,N\}, 
         u^{l} \leq u_t \leq u^{h}. \quad
    \end{aligned} \hspace{-5mm}
\end{equation}}
    
\end{definition} 

Note that the convexity of the problem assumed in Assumption \ref{assum:convex} serves the purpose of enabling the relaxation of $CCP^t_\epsilon$ to a finite number of constraints and allows for a generalization of the solution to the $CCP^t_\epsilon$ based on the solution of the simpler $RCP_N^t$. 
\begin{proposition}\label{proposition-1}
For the affine CBF  defined in 
Example E1 in Remark \ref{rem:cbf_uncertainty}, $RCP_N^t$ is convex. 
\end{proposition}

\begin{remark}\label{rem:convex}
    While we present our results for CBF constraints of the form \eqref{eq:cbf_ineq} for simplicity of exposition,  the following results are generally applicable to other forms of CBF constraints such as exponential CBFs \cite{ames2019control}, provided that they  satisfy Assumption \ref{assum:convex}. Note that Assumption \ref{assum:convex} does not hold for Example E2 in Remark \ref{rem:cbf_uncertainty} (the CBF constraint, in that case, can in fact 
be shown to be concave; see Appendix). However, it is possible to pose obstacle avoidance problems in a convex setting in certain cases using an exponential CBF formulation (one such case is presented in our case study in Section \ref{sec:case} to illustrate the broader applicability of the results in this section.)
\end{remark}

We now have the following result.

\begin{theorem} \label{thm:num_sample}
For any risk bound
 $\epsilon \in (0,1)$ and confidence parameter $\beta \in (0,1)$, if
 \vspace{-0.5mm}
 \begin{equation}\label{eq:sample_complexity}
     N \geq  \frac{2}{\epsilon}\ln{\frac{1}{\beta}} + 2m + \frac{2m}{\epsilon}\ln{\frac{1}{\beta}},
 \end{equation}
 \vspace{-0.5mm}
 then, we have that the $RCP_N^t$ is either infeasible, or if feasible, then $Prob^N\{V(u_t^*)<\epsilon\}\geq (1-\beta),$
 that is, its solution $u_t^{*}$ renders the system \eqref{dynamics} $\epsilon$-safe as in Definition \ref{def:epsilon-safe} with probability greater than or equal to 
 $1-\beta$.
\end{theorem}

Theorem \ref{thm:num_sample} provides a bound on the number of samples of the uncertainty that are required to guarantee that the control input designed by solving $RCP_N^t$ can render the system $\epsilon$-safe. 
The confidence parameter $\beta$ in \eqref{eq:sample_complexity} 
is the probability $Prob^N (= Prob \times ... \times Prob)$, (N times), of extracting  samples of the uncertainty $\{d^1_t,...,d^N_t\}$ for which the control input $u_t^{*}$ does not render the system \eqref{dynamics} safe.

We now address the question of when the scenario design problem $RCP_N^t$ for for risk-tunable control design is guaranteed to have a solution. We show that, 
under an additional assumption, $RCP_N^t$ can be shown to always be feasible.


\begin{assumption} \label{assum:cbf_bound}
 For all $d_t\in D$, there exists $u_t \in U$ such that $\bar h \in [m,M]$, $m,M\in \mathbb{R}$, $\forall x \in X$, with $M > m \geq 0$.
\end{assumption}

With this assumption, we have the following result regarding the solution of $RCP_N^t$, and therefore, the safe control design problem $CCP^t(\epsilon)$.
\begin{theorem} \label{thm:solvability}
    Let Assumptions \ref{assum:convex} and \ref{assum:cbf_bound} hold. Then, for any risk bound
 $\epsilon \in (0,1)$ and confidence parameter $\beta \in (0,1)$, if $N \geq  \frac{2}{\epsilon}\ln{\frac{1}{\beta}} + 2m + \frac{2m}{\epsilon}\ln{\frac{1}{\beta}},$
  the scenario problem $RCP^t_N$ is always solvable for any $N$ samples of the uncertainty $\{d^1_t,\ldots, d^N_t\}$, the solution $u_t^{*}$ is unique, and renders the system \eqref{dynamics} $\epsilon$-safe  in the sense of Definition \ref{def:epsilon-safe}.
\end{theorem}

Theorems \ref{thm:num_sample} and \ref{thm:solvability} provide a trade-off between the sample complexity and the achievable risk bound in the safe control design problem, representing an additional handle that can be tuned by designers based on application-specific considerations. Generally, achieving a tighter risk bound will require more samples of the uncertainty. 



\section{CASE STUDY}\label{sec:case}
\begin{figure*}
{\footnotesize
\begin{equation}\label{eq:rotation}
    \mathcal{R}_{wb}:=
    \begin{bmatrix}
    \cos{\psi}\cos\theta-\sin{\phi}\sin{\psi}\sin{\theta} & -\cos{\phi}\sin{\psi} & \cos{\psi}\sin{\theta}+\cos{\theta}\sin{\phi}\sin{\psi}\\
    \cos{\theta}\sin{\psi}+\cos{\psi}\sin{\phi}\sin{\theta} & \cos{\phi}\cos{\psi} &  \sin{\psi}\sin{\theta}-\cos{\psi}\cos{\theta}\sin{\phi}
    \\
    -\cos{\phi}\sin{\theta} & \sin{\phi} & \cos{\phi}\cos{\theta}
    \end{bmatrix}
\end{equation}}
\hrulefill
\vspace{-5mm}
\end{figure*}

We consider a quadcopter navigation problem with an obstacle whose position is uncertain to illustrate our risk-tunable design. As described in Remark \ref{rem:convex}, the CBF constraints for such obstacle avoidance problems are in general non-convex. However, in some cases, it is possible to develop convex safety conditions. We illustrate one such case here, where the nature of the dynamics arising from the system physics can be exploited to construct convex CBF conditions for the obstacle avoidance problem that are affine in the control input.

We begin with a dynamical model of the quadcopter derived in \cite{xu2018safe} and summarized here. Let the 3-dimensional position coordinates of the quadcopter along the x-,y-, and z-axis with respect to its body frame $\mathcal{F}_b$ of and the world frame of reference  $\mathcal{F}_w$ be given by $x_b:=(x_b,y_b,z_b)$ and $r:=(r_x,r_y,r_z)$ respectively. The rotation matrix for coordinate transformation from the the body frame $\mathcal{F}_b$ to the world frame  $\mathcal{F}_w$ is defined by \eqref{eq:rotation}, where $\phi$, $\theta$, and  $\psi$ denote the Z-X-Y Euler angles corresponding to the roll, pitch, and yaw of the quadcopter. 
Therefore, $r=\mathcal{R}_{wb}x_b$. Then, the quadrotor dynamics is given by 

\vspace{-3mm}
{\footnotesize\begin{align}\label{eq:quad_dynamics}
\dot x=Ax+Bu, 
  x=  \begin{bmatrix}
        \dot{r}\\
        \Ddot{r}
    \end{bmatrix}, 
  A  =
    \begin{bmatrix}
        \mathbf{0}_{3\times 3} & I\\
         \mathbf{0}_{3\times 3} & \mathbf{0}_{3\times 3}
    \end{bmatrix}, B=
    \begin{bmatrix}
        \mathbf{0}_{3\times 3}\\
        \mathbf{1}_{3\times 3}
    \end{bmatrix},
\end{align}}
\vspace{-3mm}

\noindent where the control input $u$ comprises of the desired acceleration of the quadcopter. The dynamics of the controller under small angle assumptions on the Euler angles (that is, $\sin \hat e\approx \hat e, \cos \hat e\approx 1, \hat e\in \{\phi, \theta, \psi\}$) evolves as  \cite{mellinger2012trajectory}: 

\vspace{-3mm}
{\small
\begin{align} \label{ang-to-r}
    u=
    \begin{bmatrix}
       \Ddot{r}_1^{des}\\
       \Ddot{r}_2^{des}\\
       \Ddot{r}_3^{des}
    \end{bmatrix}=
    \begin{bmatrix}  g(\theta^{des}\cos{\psi^{des}}+\phi^{des}\sin{\psi^{des}}), \\ 
   g(\theta^{des}\sin{\psi^{des}}-\phi^{des}\cos{\psi^{des}}) \\
    \frac{ \sum^4_{i=1}F_i^{des}}{m} -g \end{bmatrix},
\end{align}}
\vspace{-3mm}

\noindent where $m$ is the mass of the quadcopter, $g$ is the acceleration due to gravity, and $\Ddot{r}_i^{des}, i\in \{x,y,z\}$ is the desired acceleration component of the quadcopter in the x-,y-, and z-direction respectively, computed using the desired specifications on the Euler angles $\phi^{des}, \theta^{des}$, and $\psi^{des}$, and $F_i^{des},i\in \{1,2,3,4\}$ is the desired thrust on the $i$-th rotor of the quadcopter. 
The quadcopter dynamical parameters are set up based on \cite{hoshih2020provablyinwild}.

The objective of the control design is to enable the quadcopter to reach a target position $r_{goal}$, while avoiding an obstacle with  position $r_{obs}=\begin{bmatrix}
    r_{obs_x}+d \quad r_{obs_y}+d \quad r_{obs_z}+d
\end{bmatrix}^T$, where $d \in D\subset\mathbb{R}$ is the uncertainty in the obstacle position. For this setting, we choose a safe set $\mathcal{S}=\{r: \bar h(r)\geq 0\}$,
where
\begin{equation}\label{eq:cbf_quad}
    \bar h(r) = \left({r_{ex}}/{a}\right)^4 + \left({r_{ey}}/{b}\right)^4 + \left({r_{ez}}/{c}\right)^4 -r_s,
\end{equation}
is the CBF for system \eqref{eq:quad_dynamics}, where $r_{ex}=r_x-r_{obs_x}-d, r_{ey}=r_y-r_{obs_y}-d, r_{ez}=r_z-r_{obs_z}-d$,  with $a, b,c \in \mathbb{R}_{+}$ can be chosen to represent the shape parameters of a super-ellipsoidal obstacle, and $r_s$ is the safe distance from the obstacle to be maintained by the controller.

Now, from \eqref{eq:quad_dynamics} and \eqref{eq:cbf_quad}, notice that $\dot{\bar h}$ will not depend on the control input $u$, implying that the CBF constraint will be independent of the design variable (the control input). A standard approach to develop a CBF-based safety condition for such a system involves constructing an Exponential Control Barrier Function (ECBF) condition \cite{ames2016control} of the form  
 \begin{equation}\label{eq:ecbf}
     \Ddot{\bar h} + K\cdot[\bar h  \quad \dot{\bar h}]^T \geq 0,
 \end{equation}
where $K=[K_1 \quad K_2], K_1, K_2 \in \mathbb{R}$, is a  design parameter that  can be chosen based on the application. Now, we can rewrite \eqref{eq:ecbf} as a convex constraint in $u$ as follows. 
\begin{proposition}\label{prop:ecbf}
The barrier condition in \eqref{eq:ecbf} for the quadrotor can be setup as the affine inequality $L(x,u,d) :=-Pu-Q \leq 0$  
    with  $P{=}\begin{bmatrix}
           \frac{4r_{ex}^3}{a^4} & \frac{4r_{ey}^3}{b^4} & \frac{4r_{ez}^3}{c^4}
       \end{bmatrix}$, and  $Q=K_2 P\dot r-K_1 \bar h (r)+\dot r^T\begin{bmatrix}
          \frac{12r_{ex}^2}{a^4} & 0 &0\\
          0& \frac{12r_{ey}^2}{b^4} & 0\\
          0& 0 & \frac{12r_{ez}^2}{c^4}
       \end{bmatrix}\dot r$,
   The constraint $L(x,u,d) \leq 0$ is convex in the control input $u$. 
\end{proposition}


The formulation in Proposition \ref{prop:ecbf}  can be 
implemented in discrete-time by solving the following at each time step $t$: 
\begin{equation} \label{eq:rcp_quad}
\vspace{-0.3cm}
(RCP^{t}_N)_{q}: 
    \begin{aligned}
        \underset{u_t}{\mathrm{min}}\,  (r&-r_{goal})^T F (r-r_{goal})+ u^T G u
        \\
        \text{s.t.} \ \ L&(x_t,u_t,d^i_t) \leq 0, \ \forall i \in     \{1,...,N\}, \\
        & u^{l} \leq u_t \leq u^{h},
    \end{aligned}
\end{equation}
where $F,G\in\mathbb{R}^3$ are positive semi-definite weighting matrices. Note that $(RCP^{t}_N)_{q}$ satisfies Assumption \ref{assum:cbf_bound}. Therefore, we extract samples $\{d_i^t\}$  of the uncertain obstacle position at time $t$ according to Theorem \ref{thm:solvability}.  For our case study, the additional parameters pertaining to ${(RCP^t_N)}_q$ 
are: $d_t \in [-0.1, 0.1]$, ${r_{goal}}= (7.9, 8.1)$, ${r_{obs}}=(7.5,7.5)$, $r_s=0.4$, $(K_1,K_2)=(6,8)$, $(a,b) = 0.4$, $\beta=0.01$. The discretization time is chosen to be 0.1 sec.
Note that we only control the quadcopter along the x-
y axis, i.e. $m=2$ in Theorem \ref{thm:solvability}. 

\begin{figure}
    \centering
    \includegraphics[scale=0.43,trim=1cm 0.7cm 0.5cm 0.4cm]{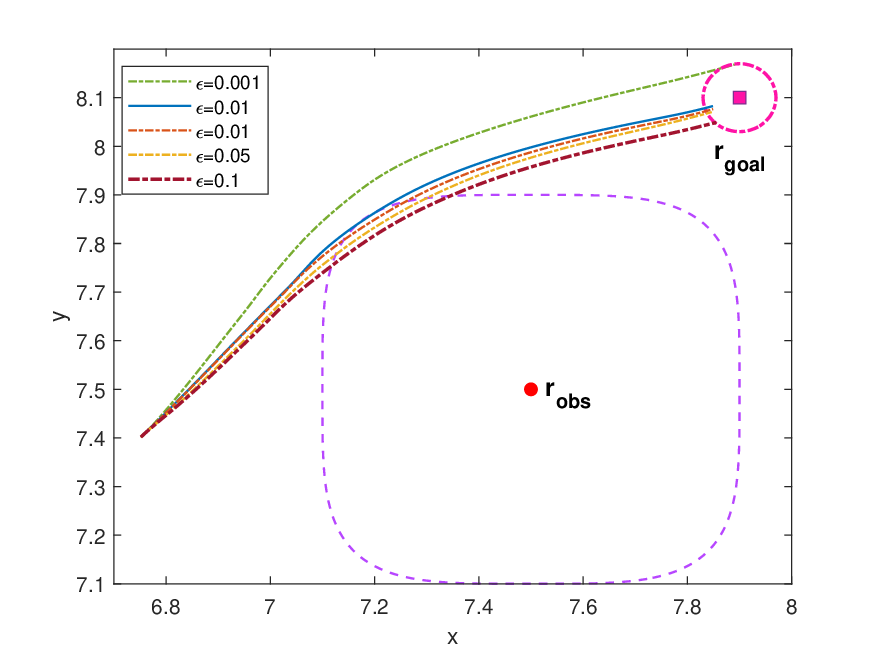}
    \caption{Impact of risk tolerance bound on controller safety and  performance.}
    \label{fig:scen_cbf_fig1}
    \vspace{-1.5em}
\end{figure}

With these parameters, we study the impact of the tunable risk tolerance $\epsilon$ on the performance of the control design. As the risk tolerance is increased (Fig. \ref{fig:scen_cbf_fig1}), the quadcopter takes a more direct path towards the goal, with some instances where it crosses into the safety margin $r_s$ around the obstacle. With a lower risk tolerance ($\epsilon=0.001$ in Fig. \ref{fig:scen_cbf_fig1}), the quadcopter takes a much longer duration to reach the goal, following a more circuitous path around the obstacle. Thus, Fig. \ref{fig:scen_cbf_fig1} illustrates how the risk in the design can be traded off for the time performance of the system.  Fig. \ref{fig:scen_cbf_fig1} also illustrates the  probabilistic nature of the safety guarantees and the role of the uncertainty in the obstacle position in this design, with the same risk  bound  $\epsilon=0.01$  resulting in two different trajectories with varying levels of safety violations.



We now examine how the sample complexity of our  design based on Theorem \ref{thm:num_sample} varies with the risk tolerance bound. Table \ref{tab:sample} lists the number of samples of the uncertainty chosen for each of the risk tolerance bounds $\epsilon$  illustrated in Fig. \ref{fig:scen_cbf_fig1}. It is observed that the sample complexity increases exponentially as the the risk tolerance bound is decreased. For this case study, we find that the risk tolerance bound $\epsilon=0.05$ provides represents an ideal design choice, bounding the risk of safety violations to under 5\%, while maintaining a reasonable trajectory to reach the goal.
\begin{table}[!h]
\centering
\caption{\label{tab:sample}Sample Complexity vs Risk Tolerance with $\beta=0.01$}
\vspace{-2mm}
\begin{tabular}{|l|l|}
\hline
Risk bound $\epsilon$ & Number of samples $N$\\ \hline \hline
0.1 & 216 \\ \hline
0.05 & 484 \\ \hline
0.01 &  3045  \\ \hline
0.001 & 39618  \\ \hline
\end{tabular}
\vspace{-1em}
\end{table}

\section{Conclusion}
We develop a safe control design approach where the probability of violation of a CBF-based safety constraint is bounded by a tunable user-defined risk, and demonstrated the design through simulations on a quadcopter navigation problem with obstacle avoidance constraints. 
Future directions include extensions to non-convex safety constraints and learning-based control designs. 



\section{Appendix}
\begin{enumerate}[leftmargin=*]
\item \textit{Proofs of Theorems  \ref{thm:cbc} and \ref{thm:cbc_uncertain}}: These results arise directly from the standard definitions of CBFs \cite{ames2019control,agrawal2017discrete}.   
\item \textit{Proof of Proposition \ref{proposition-1} and non-convexity of the CBF condition in Example E2 in Remark \ref{rem:cbf_uncertainty}}:
These proofs follow directly from the definition of convexity. 


    \item \textit{Proof of Theorem \ref{thm:num_sample}:}
    The proof is along the lines of \cite{calafiore2006scenario}.  
    We omit the dependence of all variables on time $t$ for simplicity. 
    Define $\mathcal{X}_i{:=}\{u\in U: L(x,u,d^i)\leq0\}$.
 From Assumption \ref{assum:convex}, $\{\mathcal{X}_i\}$, $\forall i \in [1,N+2m]$ are the convex sets defined by the $2m$ actuator constraints  $u\in U=[u^l,u^h]$ and the $N$ CBF constraints in $RCP_N^t$. Define convex optimization problems $\mathcal{P}$  and $\mathcal{P}_k$, $k{\in}[1,N+2m]$, obtained by removing the $k^{th}$ constraint as:

\vspace{-2mm}
 {\footnotesize
    \begin{equation*}
          \begin{aligned}
           &\mathcal{P}:   \underset{u_t}{\mathrm{min}}\ C(u_t), \quad 
            s.t. \ u_t{\in}\bigcap_{i=\{1,...,N+2m\}}\mathcal{X}_i, \\
    &\mathcal{P}_k:
            \underset{u_t}{\mathrm{min}}\ C(u_t) , \quad 
             s.t. \ u_t \in \bigcap_{i=\{1,...,N+2m\}\setminus k}\mathcal{X}_i
        \end{aligned} 
    \end{equation*} }
\vspace{-3mm}

 Suppose $RCP_N^t$ is feasible, and $u_t^*$ is the optimal solution to $\mathcal{P}$, and $u_k^*$ is the optimal solution to $\mathcal{P}_k$. Then, the $k^{th}$ constraint is a \textit{support constraint} if $C(u^*_k){<}C(u^*)$. 
 The number of support constraints for problem $\mathcal{P}$ is at most $m$ \cite[Theorem 3]{calafiore2006scenario}. 
Given $N$ scenarios $\{d^1,...,d^N\}$, select a subset $\mathcal{I} = \{i_1,...,i_{m}\}$ of $m$ indices from
$1,...,N+2m$ and let $\hat{u}_I^{*}$ be the optimal solution of ${RCP_{I}^t}$ defined as 

\vspace{-1.5mm}
{\footnotesize
\begin{equation}
    RCP_{I}^t:
    \begin{aligned}
        &\underset{u_t}{\mathrm{min}}\   C(u_t)
        \\
        &s.t.  \  L(x,u,d^i) \leq 0, \ \forall i \in     \{1,...,m\}, \\
        & u^{l} \leq u \leq u^{h}.
    \end{aligned}
\end{equation}}
\vspace{-4mm}

Let $\Delta^N{:=}\{d^i \}_{i=1,...,N}$ be the set of all possible $N$ samples drawn from  set $D$. Define  
$\Delta^N_I{\subset}\Delta^N$  as $\Delta^N_I{=}\{d^1,...,d^N{:}\hat{u}_I^{*}{=}\hat{u}_N^{*}\}$
where $\hat{u}_N^{{*}}$ is the optimal solution with all $N$ constraints corresponding to $\{d^1,...,d^N\}$.
Let $\mathcal{I}$ be a collection of all possible choices of $m$ indices from $1,...,N+2m$, then $\mathcal{I}$ contains $\begin{pmatrix}
    N+m \\ m
\end{pmatrix}$ sets and $\Delta^N{=}\bigcup_{I\in \mathcal{I}}\Delta_{I}^N.$
Now suppose,
    $B{:=}\{d^1,...,d^N:V(\hat{u}_N^{*}){>}\epsilon\}$
and
    $B_I{:=}\{d^1,...,d^N:V(\hat{u}_I^{*}){>}\epsilon\}$.
Then, $B=\bigcup_{I\in\mathcal{I}}(B_I \bigcap\Delta^N_I).$
A bound for $Prob^N(B)$ is now obtained by bounding $Prob(B_I\bigcap \Delta_I^N)$ and then summing
over $I{\in}\mathcal{I}$. Following a similar argument as in the proof in \cite[Appendix B]{calafiore2006scenario}, we have $Prob^N(B){\leq}\sum_{I\in \mathcal{I}}Prob^N(B_I\bigcap\Delta^N_I)$
which can further be bounded as $\sum_{I\in \mathcal{I}}Prob^N(B_I\bigcap\Delta^N_I){<}\begin{pmatrix}
    N+m \\ m
\end{pmatrix}(1{-}\epsilon)^{N+m},$ since 
$\mathcal{I}$ has  $\begin{pmatrix}
    N+m \\ m
\end{pmatrix}$ sets.  
Then, following the algebraic manipulations in \cite[Appendix B]{calafiore2006scenario}, from \eqref{eq:sample_complexity}, 
we have $\begin{pmatrix}
    N+m \\ m
\end{pmatrix}(1-\epsilon)^{N+m} \leq \beta$, that is, $Prob^N(B) < \beta.$
Now, if $RCP_N^t$ is only feasible on a subset $F_s{\subset}\Delta^N$, the same arguments hold to prove that $Prob^N(B){<}\beta.$ holds in the set $F_s$, with $B{:=}\{(d^1,...,d^N){\in}F_s:V(\hat{u}_N^{*}) > \epsilon\}$. 

    \item \textit{Proof of Theorem \ref{thm:solvability}}: Since $h\in [m,M]$, $\forall u \in [u^{l},u^{h}]$, we have $ -L(x_t,u_t,d_t) \geq m-(1-\eta)M, \ \forall d \in D,  \forall t \in \mathbb{R}_{+}.$      Then, we can always select $\eta$, such that  $1 \geq \eta \geq 1-\frac{m}{M}$ to ensure $L(x_t,u_t,d_t) \leq 0.$

      \item \textit{Proof of Proposition \ref{prop:ecbf}}: 
      The proof follows from  differentiating \eqref{eq:cbf_quad}, substituting in \eqref{eq:ecbf}, collecting the terms with the control input $u$, and invoking Proposition \eqref{proposition-1}.
\end{enumerate}

\balance
\bibliographystyle{IEEEtran}
\bibliography{references}

\end{document}